# Broadcast Approach and Oblivious Cooperative Strategies for the Wireless Relay Channel – Part II : Block-Markov Decode-and-Forward (BMDF)

Evgeniy Braginskiy , Avi Steiner and Shlomo Shamai (Shitz)

**Abstract**

This is the second in a two part series of papers on incorporation of the broadcast approach into oblivious protocols for the relay channel where the source and the relay are collocated. Part I described the broadcast approach and its benefits in terms of achievable rates when used with the sequential decode-and-forward (SDF) scheme. Part II investigates yet another oblivious scheme, the Block-Markov decode-and-forward (BMDF) under the single and two-layered transmissions. For the single layer, previously reported results are enhanced and a conjecture regarding the optimal correlation coefficient between the source and the relay's transmission is established. For the discrete multi-layer transmission of two or more layers, it is shown that perfect cooperation (2x1 MISO) rates are attained even with low collocation gains at the expense of a longer delay, improving upon those achievable by the SDF.

I. INTRODUCTION

In part I of this two-part series [1] , we introduced the incorporation of the broadcast approach into an oblivious cooperation protocol termed sequential decode-and-forward (SDF) which is used to enhance the achievable rates for a source communicating to a destination with an aid of a nearby relay. It was demonstrated that by using two-layered transmission, substantial gains can be achieved under various channel settings, for both oblivious and non-oblivious cooperation scenarios ,while enjoying low delay. The two-layered transmission achieved about 80% of the optimal continuous layering gain, with further improvement possible by increasing the number of layers.

The main limitation of the SDF is the fact that the relay remains silent while trying to decode the message, which is rather a severe constraint. Had the relay been unable to decode the message during the block time, the destination views a single user channel with its low achievable rates. The ability of the two-layer approach to overcome this drawback is partial , mainly due to its interference limited nature. In order to

E.Braginskiy , A.Steiner and S.Shamai (Shitz) are with the Department of Electrical Engineering, Technion- IIT, Haifa , 32000, Israel. Email {bevgeniy@tx, savi@tx, sshlomo@ee}.technion.ac.il. This work was supported by the Israel Science Foundation and by the European Commission in the framework of the FP7 Network of Excellence in Wireless Communications NEWCOM++.

alleviate this constraint, in part II we consider the Block-Markov encoding scheme which originates from [3]. We employ the regular encoding/backward decoding scheme [6] at the destination for the two-layer transmission to show that very significant gains are achieved over the direct transmission even with low collocation gains. The case of correlated transmission of the source and the relay is examined [2], [4] for both the outage and the broadcast approaches and the optimality of the uncorrelated scheme $(\rho=0)$ is established for the outage approach. In [4], the maximal throughput for a 2x1 MISO is established analytically under individual power constraints and constrained correlation, proving useful in a BM rate evaluation. Another result there [Theorem 1] is used to reduce the optimal correlation uncertainty area of [2, Theorem 1] under equal source and relay powers.

The rest of the paper is organized as follows. Section II gives the necessary background to the BM scheme; Section III discusses the single layer BM; Section IV treats the embedding of a two-layer broadcasting into the BM scheme; finally concluding remarks are given in Section V. Proofs and derivations are deferred to Appendixes A-B.

## II. THE BLOCK-MARKOV ENCODING/DECODING SCHEME

In order to set up the discussions in this paper, we reiterate the channel model used in part I which is illustrated in Fig. 1 and can be mathematically expressed as

$$Y_r(n) = \sqrt{Q} X_s(n) + Z_r(n)$$
$$Y_d(n) = h_s X_s(n) + h_r X_r(n) + Z_d(n) \quad . \quad (1)$$

Otherwise, the signal at the destination is modeled by

$$Y_d(n) = h_s X_s(n) + Z_d(n). \quad (2)$$

In (1), (2), $X_s(n)$ and $X_r(n)$ are the symbols transmitted during the $n$-th symbol interval by the source and the relay respectively, $Z_r(n)$ and $Z_d(n)$ are the AWGN at the relay and the destination respectively, both modeled by i.i.d complex Gaussian r.v's with zero mean and unit variance, $Q$ is the collocation gain and $h_s, h_r$ are the Rayleigh fading coefficients for the source-destination and relay-destination channels. Further details on this model can be found in part I.

The decode-and-forward strategy for the relay channel [3, Theorem 1] achieves any rate up to

$$I_{DF,CE} = \max_{p(x_s, x_r)} \min \{I(x_s : y_r | x_r), I(x_s, x_r : y_d)\} \quad (3)$$

and is actually the capacity of a degraded relay channel [3]. Expression (3) can be interpreted as the minimum between the maximal rate supported by the source-relay channel and the maximal rate supported by the 2x1 MISO channel consisting of the source/relay and the destination. There are three known schemes which achieve the rate (3). The first scheme, irregular encoding/successive decoding, developed in [3], involved Markov superposition coding, random binning and successive decoding. The encoding of the source and the relay messages was done with codebooks of different size, hence the term irregular encoding. The scheme of the regular encoding/sliding-window decoding appears in a work of Carleial [7] in the context of multiple-access channel with generalized feedback (MAC-GF) in which the destination employed a sliding-window decoding by using two consecutive outputs of the channel and the coding was done with codebooks of equal sizes. The MAC-GF of Carleial includes the relay channel as a special case. The third strategy , regular encoding/backward decoding , was developed by Willems [6] in his work on the MAC-GF. The backward decoding technique is generally better than the sliding-window decoding, but for the discrete memoryless relay channel the three schemes achieve the same rate. The structure of the superposition code for the fading channel as well as the encoding/decoding procedure can be found in [9].

### III. SINGLE LAYER BLOCK-MARKOV DECODE-AND-FORWARD

We start by examining the single layer case. The relay and the destination signals are defined in (1) and a correlation between the source and the relay signals is introduced by defining

$$\rho \triangleq \frac{E[X_s X_r^*]}{\sqrt{P_s P_r}}.$$

The DF rate for the relay channel becomes [2]

$$I_{DF,CE} = \min_{p(x_s,x_r)} \left[ \log\left(1 + P_s Q \left(1 - |\rho|^2\right)\right), \log\left(1 + v_s P_s + v_r P_r + 2\sqrt{P_s P_r} \Re_e \left(\rho h_s h_r^*\right)\right) \right] \quad (4)$$

The solution to (4) is given by [2, Theorem 1], from which three rate regions can be defined : low rates with $\rho^{opt} = 0$, intermediate rates with either $\rho^{opt} = 0$ or $\rho^{opt} = \rho_{\max}$ and high rates with $\rho^{opt} = \rho_{\max}$. Increasing the power asymmetry between the source and the relay diminishes the intermediate region. Examination of the BM rates as compared to the cut-set bound [5] reveals equality for certain channel conditions, especially when $Q$ is high, expressing a proximity between the source and the relay. This means the BM rate is the capacity under those conditions in agreement with the fact that DF is optimal when the relay is near the source [5]. As $Q \to \infty$ we may choose $\rho = 1$ and remain with the second term only, which is a cut rate and therefore optimal. In this case, $I(x_s, x_r : y_d)$ becomes the limiting term. Another observation from the various settings examined in this work is that the highest throughput is attained with $\rho = 0$. This is stated in general in the following conjecture.

*Conjecture 1*: For a relay channel described by (1) using the BM protocol, the highest average throughput is attained for $\rho = 0$.

Conjecture 1 is suggested by the derivations of Appendix A coupled with the numerical results.

As stated in [2, Theorem 1], in the rate region $R \in \left[\log(1+P), \log(1+\gamma P)\right], \gamma \leq \frac{3}{2}$ the optimal $\rho$ is unknown and may be either of the two options depending on the parameters. For $P_s = P_r$, we have $\gamma = \frac{3}{2}$. However, for equal powered source and relay, and sufficiently high $Q$, the uncertainty region may be reduced by using the following theorem.

*Theorem 1*: For a relay channel described by (1) using the BM protocol, the region in which $\rho^{opt} = 0$ or $\rho^{opt} = \rho_{\max}$ is

$$R \in \left[R_0 \triangleq \log(1+\gamma_0 P), \log(1+\gamma P)\right]$$

$$\gamma_0 = -W_{-1}\left(-\frac{1}{2}e^{-\frac{1}{2}}\right) - \frac{1}{2} \cong 1.2564$$

where $W_{-1}(x)$ is the -1-th branch of the Lambert W function [8] defined for $-\frac{1}{e} < x < 0$ as the solution to the equation $W_{-1}(x)e^{W_{-1}(x)} = x$ satisfying $W_{-1}(x) < -1$. In addition, if $Q \to \infty$, then

$$R < \log(1+\gamma_0 P), \rho^{opt} = 0, R > \log(1+\gamma_0 P), \rho^{opt} = \rho_{\max} = 1$$

*Proof*: Suppose $P_s = P_r$, making the total power budget of the system constrained by $P = 2P_s$. Next, let us compare the decoding probability of a single user with a power constraint of $P$ to the decoding probability of the source and the relay cooperating as a 2x1 MISO with equal powers of the terminals. For the single user and the MISO respectively

$$P^{SU}(R) = e^{-\frac{e^R-1}{2P_s}}, P^{MISO}(R) = \left(1 + \frac{e^R-1}{P_s}\right)e^{-\frac{e^R-1}{P_s}}.$$

Defining $x \triangleq \frac{e^R-1}{P_s}$, the probabilities are equal when

$$(1+x)e^{-\frac{x}{2}} - 1 = 0 \to -\left(\frac{x+1}{2}\right)e^{-\left(\frac{x+1}{2}\right)} = -\frac{1}{2}e^{-\frac{1}{2}}$$

and the solution is given in terms of the -1-th branch of the Lambert W function as

$$x_0 = -2W_{-1}\left(-\frac{1}{2}e^{-\frac{1}{2}}\right) - 1 \cong 2.5128$$

meaning that for $R < \log(1+x_0 P_s)$ the MISO setting is superior to that of a single user and vice versa, in agreement with [5, Theorem 1], by which the decoding probability is maximized under equal power splitting between the terminals for $R < R_0$. As this is achieved with $\rho = 0$, $\rho^{opt} = 0, R < R_0$. For $R > R_0$ the achievable rate with $\rho = 0$ is no longer necessary optimal. Thus, other correlation coefficients may lead to higher rates by a more sophisticated power splitting. But, from [4, Theorem 2], only two values of correlation coefficients are of interest for a correlation constrained MISO, therefore for the rates satisfying $R \in [\log(1+\gamma_0 P), \log(1+\gamma P)]$ both options remain valid, completing the first step of the proof.

Next, the average throughput with $\rho_{max}$ as the correlation coefficient is given by

$$R_{av}^{COR}(R) = r\left(P_0, \overline{P_0}, e^R - 1\right), P_0 = \frac{1}{2}\left[P + \sqrt{P^2 - 4P_r\left(\frac{e^R - 1}{Q}\right)}\right], \overline{P_0} = P - P_0.$$

The effect of the correlation can be seen as a re-allocation of the available power between the source and the relay according to

$$P_s^{new} = \left(\frac{P_s + P_r}{2}\right)(1+\delta), P_s^{new} = \left(\frac{P_s + P_r}{2}\right)(1-\delta), 0 \leq \delta \leq 1,$$

with the skew $\delta$ increasing with the collocation gain $Q$. For $Q \to \infty$ the uncertainty region vanishes and the rate $R_0$ is the boundary between $\rho^{opt} = 0$ and $\rho^{opt} = \rho_{max}$ since the single user and the correlated MISO coincide for all decodable rates by $P_0 \to P\square$.

Fig. 2 displays an application of Thm. 1 for finite collocation gain, where even for a moderate $Q$ of 10dB the difference between $R_0$ and the first attempted rate for which $\rho^{opt} \neq 0$ is optimal is about 0.02 [nats/channel use], which is practically insignificant.

Next, we proceed to evaluating the achievable rates for the BM scheme under the oblivious setting. Due to the ability of achieving the MISO rates even for finite collocation gains as mentioned, the rates are expected to be independent of $Q$ for $Q > Q_{min}$ defined in (5). In Fig. 3, the oblivious rates are presented for several source-relay channel qualities. The BM scheme exhibits significant gains over the direct transmission, and, more importantly, under those channel conditions the gains are independent of the collocation gain. This is since MISO term of (3) is the limiting factor of the achievable rates. The gain is about 4dB over the given $P_s$ range, approximately 1dB higher than for the SDF, as MISO rates are achieved here for finite $Q$ in contrast to the SDF which required $Q \to \infty$, provided $R < \log(1 + P_s Q)$. Note that for the oblivious rates, the condition for the BM to achieve MISO performance is

$$R_{opt,SU} = W(P_s) < \log(1 + P_s Q)$$

and since $W(P_s) < \log(1+P_s)$ with the difference increasing with $P_s$, the condition will hold for $P_s$ high enough. For a given $P_s$, the minimal $Q$ required to allow MISO rates under oblivious cooperation is given by

$$Q_{min} = \frac{e^{W(P_s)} - 1}{P_s} = \frac{1}{W(P_s)} - \frac{1}{P_s} < 1 \tag{5}$$

and $Q_{min}$ is a decreasing function since

$$\frac{\partial Q_{\min}}{\partial P_s} = \frac{-W(P_s)'}{W(P_s)^2} + \frac{1}{P_s^2} = -\frac{1}{W(P_s)(1+W(P_s))P_s} + \frac{1}{P_s^2} < 0, \frac{P_s}{W(P_s)} = e^{W(P_s)} > 1 + W(P_s), \lim_{P_s \to \infty} Q_{\min} = 0,$$

thus an achievability of the MISO rates for the single layer BM is established. Expression (5) is plotted in Fig. 4 for a wide range of source powers with the transition from the direct transmission occurring at a certain power threshold depending on $Q$. From (5) we can analytically derive the threshold $P_s^*$ defined as the minimal $P_s$ for a given $Q$ such that for $P_s > P_s^*$ the MISO rates are achievable. The following proposition defines $P_s^*$.

*Proposition 1*: For a given collocation gain $Q < 1$, the MISO rates are achievable for

$$P_s > P_s^* \triangleq -\frac{1}{Q}\left(1 - e^{W\left(-\frac{1}{Q}e^{-\frac{1}{Q}}\right) + \frac{1}{Q}}\right).$$

*Proof:* For $P_s^*$ we have that $W(P_s^*) = \log(1 + P_s^* Q)$, therefore the following equality takes place

$$W(P_s^*) e^{W(P_s^*)} = P_s^* \Rightarrow \left[\log(1 + P_s^* Q) - \frac{1}{Q}\right](1 + P_s^* Q) = -\frac{1}{Q}.$$

Multiplying both sides by $e^{-\frac{1}{Q}}$ and rearranging, $P_s^*$ satisfies

$$\left[\log(1 + P_s^* Q) - \frac{1}{Q}\right] e^{\left[\log(1 + P_s^* Q) - \frac{1}{Q}\right]} = -\frac{1}{Q} e^{-\frac{1}{Q}},$$

resulting in

$$P_s^* = -\frac{1}{Q}\left(1 - e^{W\left(-\frac{1}{Q}e^{-\frac{1}{Q}}\right) + \frac{1}{Q}}\right) \square.$$

Solving for $Q > 1$ requires the use of $W_{-1}(x)$ and although the solution is similar, the resulting $P_s^*$ is negative, indicating that the MISO rates are attained unconditionally for positive (in dB) collocation gains.

## IV. Two layer broadcasting BM

### A. Uncorrelated transmission rates

In this section, we incorporate the two layer broadcasting into the BM structure. The most appealing feature of the BM as witnessed from the numerical results it that the collocation gain parameter does not need to be infinite to achieve the MISO bound. We therefore expect the two layer BM to achieve the MISO broadcasting bound for appropriate $Q$ under the oblivious cooperation. For the sake of mathematical tractability, we will assume that the relay assists only when decoding both layers of the fractional message.

We use the same layering rate definitions as for the SDF and encode each layer with a BM code. The destination applies successive decoding to the layered transmission. The average throughput for each of the layers can be written as

$$R_{av}^{BM} = R \cdot P(I_{DF,CE} > R) = R \cdot P(\min(A,B) > R) = R \cdot P((A>R) \cap (B>R)), A = I(x_s : y_r | x_r), B = I(x_s, x_r : y_d) \quad (6)$$

where $R = R_1, R_2$ respectively. In our model, the source-relay channel is gain fixed, therefore for an attempted rate $P(A>R)$ is an indicator function and by taking $Q$ large enough compared to $\eta_1, \eta_2$, a reliable transmission over the source-relay link can be achieved. We start with deriving the expressions $A, B$ for the two layer transmission bearing in mind the successive decoding of the layers at the destination coupled with backwards decoding strategy. Starting with the general system model, rewrite it as

$$y_r = \sqrt{Q} X_s + Z_r = \sqrt{Q} X_{s,1} + \sqrt{Q} X_{s,2} + Z_r$$
$$y_d = h_s X_s + h_r X_r + Z_d = (h_s X_{s,1} + h_r X_{r,1}) + (h_s X_{s,2} + h_r X_{r,2}) + Z_d$$

where the transmitted signal is split into two layers, emphasizing the additive correlated interference added to the first layer by the second. We consider variable power allocation and similarly to the single layer case define two correlation coefficients by

$$\rho_1 \triangleq \frac{E[X_{s,1} X_{r,1}^*]}{\sqrt{\alpha \beta P_s P_r}}, \rho_1 \triangleq \frac{E[X_{s,2} X_{r,2}^*]}{\sqrt{\bar{\alpha}\bar{\beta} P_s P_r}}. \quad (7)$$

The following theorem states the achievable average throughput for the two layer Block-Markov protocol.

*Theorem 2*: For a relay channel described by (1) using the BM protocol with two layer broadcasting, the average achievable throughput is given by

$$R_{av}^{BM,B} = R_1 \cdot P(R_1^P \cap (R_2^P)^c) + (R_1 + R_2) \cdot P(R_1^P \cap R_2^P) = R_1 \cdot P(R_1^P) + R_2 \cdot P(R_1^P \cap R_2^P)$$

$$R_1^P \triangleq \left( \log \left( \frac{1 + P_s Q (1 - \alpha\beta|\rho_1|^2 - \bar{\alpha}\bar{\beta}|\rho_2|^2 - 2\sqrt{\alpha\beta\bar{\alpha}\bar{\beta}} \Re_e(\rho_1 \rho_2^*))}{1 + Q \bar{\alpha} P_s \cdot (1 - |\rho_2|^2)} \right) > R_1 \right) \cap \left( \log \left( \frac{1 + v_s P_s + v_r P_r + 2(\sqrt{\alpha\beta P_s P_r} \Re_e(\rho_1 h_s h_r^*) + \sqrt{\bar{\alpha}\bar{\beta} P_s P_r} \Re_e(\rho_2 h_s h_r^*))}{1 + v_s \bar{\alpha} P_s + v_r \bar{\beta} P_r + 2 \sqrt{\bar{\alpha}\bar{\beta} P_s P_r} \Re_e(\rho_2 h_s h_r^*)} \right) > R_1 \right)$$

$$R_2^P \triangleq \left( \log(1 + Q \bar{\alpha} P_s (1 - |\rho_2|^2)) > R_2 \right) \cap \left( \log(1 + v_s \bar{\alpha} P_s + v_r \bar{\beta} P_r + 2 \sqrt{\bar{\alpha}\bar{\beta} P_s P_r} \Re_e(\rho_2 h_s h_r^*)) > R_2 \right)$$

(8)

*Proof*: Appendix B.

Examination of the decoding probabilities in (8) reveals that by taking $\rho_1 = \rho_2 = 0$ we get the same quantities for the mutual information as in the variable MISO case for the MISO term while the source-relay expression defines the rates decodable by the relay. By the layered rate definitions it is readily seen that choosing $Q > \eta_2$ ensures the source-relay channel supporting the attempted rate. Using $\rho_1 = \rho_2 = 0$ allows for highest decodable rates over the source-relay channel, and the expressions for the second layer are the same as in the single layer case up to a power scaling, due to the removed interference from the first layer assumed in its decoding.

As the phase of $h_s h_r^*$ is uniform over $[0, 2\pi)$, we can restrict ourselves to non-negative and real values for the correlation coefficients. For a pair of rates $R_1, R_2$, $\rho_2$ has to be constrained to the interval $\left[0, \sqrt{1 - \frac{e^{R_2} - 1}{Q \bar{\alpha} P_s}} = \sqrt{1 - \frac{\eta_2}{Q}}\right]$ for the first term of $R_2^p$ (8) to support the rate. Under the mentioned conditions, the average throughput can be computed as

$$R_{av}^{BM,B} = R_1 \cdot P\left(\begin{array}{l}\left(\log\left(\frac{1 + v_s P_s + v_r P_r}{1 + v_s \bar{\alpha} P_s + v_r \bar{\beta} P_r}\right) > R_1\right) \\ \cap \left(\log\left(1 + v_s \bar{\alpha} P_s + v_r \bar{\beta} P_r\right) < R_2\right)\end{array}\right) + (R_1 + R_2) \cdot P\left(\begin{array}{l}\left(\log\left(\frac{1 + v_s P_s + v_r P_r}{1 + v_s \bar{\alpha} P_s + v_r \bar{\beta} P_r}\right) > R_1\right) \\ \cap \left(\log\left(1 + v_s \bar{\alpha} P_s + v_r \bar{\beta} P_r\right) > R_2\right)\end{array}\right). \quad (9)$$

with (9) already solved for the variable power allocation MISO under the SDF setting [1] with the difference being no restriction on the source-relay rates as the channel allowed perfect cooperation. Fig. 5 demonstrates the average throughput of the two layer BM with uncorrelated transmission for various collocation gains. Under those conditions BM achieves the MISO broadcasting bond with a very substantial gain of about 4.5-5dB over the direct transmission even for the oblivious setting limitations. The gains are about 0.5dB larger than those achieved by the BM for a single layer, and the gain of the two layer rate over its single layer counterpart is about 2-3dB.

From , much higher $P_s$ values are required for the BM to approximate the MISO bound in the two layered case than in the single layer case. This can be attributed to the lower bound on $Q$ behaving differently for each of the methods, with a weaker dependence on $P_s$ for broadcasting. The two layer broadcasting lower bound on $Q$ is given by

$$Q_{\min,2} > \eta_2^{opt} = \frac{e^{W(\bar{\alpha}^{opt} P_s)} - 1}{\bar{\alpha}^{opt} P_s} = \frac{1}{W(\bar{\alpha}^{opt} P_s)} - \frac{1}{\bar{\alpha}^{opt} P_s} \quad (10)$$

and by differentiation, we get

$$r \triangleq \bar{\alpha}^{opt} P_s,$$

$$\frac{\partial Q_{\min,2}}{\partial P_s} = \frac{\partial Q_{\min,2}}{\partial r} \cdot \frac{\partial r}{\partial P_s} = \frac{\partial Q_{\min,2}}{\partial r}\left(\bar{\alpha}^{opt} + P_s \left(\bar{\alpha}^{opt}\right)'\right) = \underbrace{\left(-\frac{1}{W(r)(1 + W(r))r} + \frac{1}{r^2}\right)}_{<0}\left(\bar{\alpha}^{opt} + P_s \left(\bar{\alpha}^{opt}\right)'\right)$$

and assuming $Q_{\min,2}$ is a decreasing function, the rate of decrease is expected to be lower than for the single layer, and $Q_{\min,2} \geq Q_{\min}$, as shown in. Practically, this means that for a low collocation gain, we can expect to achieve the MISO bound only for a very large $P_s$, if any. By Proposition 1, the threshold for the two layer case satisfies

$$\bar{\alpha}_{opt}\left(P_{s,B}^{*}\right)P_{s,B}^{*} = -\frac{1}{Q}\left(1 - e^{W\left(-\frac{1}{Q}e^{-\frac{1}{Q}}\right) + \frac{1}{Q}}\right)$$

and obviously $P_{s,B}^{*} \geq P_s^{*}$, resulting in non-practically large source powers even for a moderate $Q < 0\text{dB}$. Numerical methods indicate that $\bar{\alpha}_{opt}(P_s)P_s$ is a monotonically increasing unbounded function. Finally, Fig. 6 demonstrates the rates when the relay power is fixed. Significant gains are achieved, decreasing with the increase of source power as relay's contribution is shadowed by the source dominance with increasing $\frac{P_s}{P_r}$. The MISO rates are unconditionally achieved for such $Q$.

A further improvement of the rates is possible by considering multi-layer transmission with $N > 2$ [1]. In this case, the SISO attempted rates are chosen to maximize $R_{av} = \sum_{i=1}^{N} R_i e^{-\eta_i}$. For the equal power allocation MISO, those can then be plugged into [1, sec V.C] to compute the rate achievable by the BM, provided that $Q > \eta_N$. Although higher $Q$'s will be required since $\eta_N > \eta_2, N > 2$, still $Q > 1$ ensures achievement of MISO rates since $\frac{1}{W(x)} - \frac{1}{x} < 1, \forall x > 0$.

*B. Correlated transmission*

Having examined the $\rho_1 = \rho_2 = 0$ MISO case, we turn to evaluating the general expressions. Rearranging the left-hand term of (39) and restricting the correlation values to be to real and positive, one gets

$$\mathbf{1}(\rho_1, \rho_2) \triangleq \rho_1^2\left(-P_s Q\alpha\beta\right) + \rho_2^2\left(Q\bar{\alpha}P_s e^{R_1} - P_s Q\bar{\alpha}\bar{\beta}\right) + \rho_1\rho_2\left(-2P_s Q\sqrt{\alpha\beta\bar{\alpha}\bar{\beta}}\right) + \left(1 + P_s Q - e^{R_1}\left(1 + Q\bar{\alpha}P_s\right)\right) > 0 \quad (11)$$

Recall that the general quadratic equation in two variables is a conic section assuming a form of $Ax^2 + Bxy + Cy^2 + Dx + Ey + F = 0$, and since $B^2 - 4AC > 0$ in our case, (11) represents an area constrained between the hyperbola branches for $F > 0$ and outside the branches for $F < 0$. Therefore, the intersection of this area with the $[0,1] \times [0,1]$ rectangle determines the feasible correlation coefficients for a given attempted rate. If the intersection is void, the first layer is un-decodable, nullifying the throughput. Note that $F > 0$ ensures that the first layer is decodable at least for $\rho_1 = \rho_2 = 0$ if $Q > \eta_1$.

For a fixed $\rho_2$, the LHS of (11) is maximal for $\rho_1 = 0$ since the function has a negative derivative w.r.t $\rho_1$. For a fixed $\rho_1$, on the other hand, the derivative becomes $2P_s Q\bar{\alpha}\rho_2\left(e^{R_1} - \bar{\beta}\right) - 2P_s Q\rho_1\sqrt{\alpha\beta\bar{\alpha}\bar{\beta}}$ and it is a linear function of $\rho_2$. The function itself therefore starts by decreasing from its value at $(\rho_1, 0)$ (unless $\rho_1 = 0$) and then either reaches a minimum and increases from it or decreases monotonically up to the point $(\rho_1, 1)$. In both cases, the maximum will be attained at the boundary, depending on the parameters. By setting $\rho_1 = 0$, the constrained maximum is attained with $\rho_1 = 0, \rho_2 = 1$ by positivity of the stated derivative, and defines the maximal decodable first layer rate by

$$R_1 \leq \log\left(1 + P_s Q\left(1 - \bar{\alpha}\bar{\beta}\right)\right). \quad (12)$$

Attempting higher rate nullifies throughput as well since the second layer is not supported for $\rho_2 = 1$. Examination of the second term of (39) reveals as expected that decreasing $\rho_1$ also decreases the SNR over the MISO channel, thus presenting a tradeoff for the optimal $\rho_1$ similarly to the single layer expressions. Since the optimization is on the overall throughput, the optimality of the single layer solution for the second layer is no longer necessary valid, as using low $\rho_2$ results in higher interference to the first layer while increasing $\rho_2$ reduces the second layer rate. Some intermediate values of $\rho_2$ can therefore be expected.

Let us now derive the explicit probability for the second term of (39). First, notice that

$$P\left(I\left(x_{s,1}, x_{r,1} : y_d\right) > R\right) = \int_0^\infty P\left(I\left(x_{s,1}, x_{r,1} : y_d\right) > R \mid v_s = a\right) e^{-a} da$$

$$P\left(I\left(x_{s,1}, x_{r,1} : y_d\right) > R \mid v_s = a\right) = P\left(\log\left(\frac{1 + aP_s + v_r P_r + 2\left(\rho_1\sqrt{\alpha\beta P_s P_r a}\Re_e(h_r) + \rho_2\sqrt{\bar{\alpha}\bar{\beta}P_s P_r a}\Re_e(h_r)\right)}{1 + a\bar{\alpha}P_s + v_r \bar{\beta}P_r + 2\rho_2\sqrt{\bar{\alpha}\bar{\beta}P_s P_r a}\Re_e(h_r)}\right) > R\right) \quad (13)$$

and by rewriting $h_r = ue^{j\phi}$ (13) is given by

$$P\left(\log\left(\frac{1 + aP_s + u^2 P_r + 2\left(\rho_1\sqrt{\alpha\beta P_s P_r}au\cos(\phi) + \rho_2\sqrt{\bar{\alpha}\bar{\beta}P_s P_r}au\cos(\phi)\right)}{1 + a\bar{\alpha}P_s + u^2\bar{\beta}P_r + 2\rho_2\sqrt{\bar{\alpha}\bar{\beta}P_s P_r}au\cos(\phi)}\right) > R\right) = \begin{cases} \int_0^\infty 2ue^{-u^2} P(\cos(\phi) > f(u)) du, f(u) = \frac{e^R\left(1 + a\bar{\alpha}P_s + u^2\bar{\beta}P_r\right) - 1 - aP_s - u^2 P_r}{u\sqrt{a}\left(2\rho_1\sqrt{\alpha\beta P_s P_r} + 2\rho_2\left(1 - e^R\right)\sqrt{\bar{\alpha}\bar{\beta}P_s P_r}\right)} > 0 \\ \int_0^\infty 2ue^{-u^2} P(\cos(\phi) < f(u)) du, f(u) = \frac{e^R\left(1 + a\bar{\alpha}P_s + u^2\bar{\beta}P_r\right) - 1 - aP_s - u^2 P_r}{u\sqrt{a}\left(2\rho_1\sqrt{\alpha\beta P_s P_r} + 2\rho_2\left(1 - e^R\right)\sqrt{\bar{\alpha}\bar{\beta}P_s P_r}\right)} < 0 \end{cases}$$

(14).

where the cumulative distribution of $\cos(\phi), \phi \sim U[0, 2\pi]$ is given by [5] as

$$P(\cos(\phi) < f(u)) = \begin{cases} 1, f(u) > 1 \\ \frac{1}{2} + \frac{1}{\pi}\arcsin(f(u)), -1 \leq f(u) \leq 1 \\ 0, f(u) < -1 \end{cases} \quad (15).$$

We now derive a more explicit expression for $P(\cos(\phi) > f(u))$ depending on the value of $a$. In general, $f(u) > 1$ implies

$$k \triangleq \sqrt{a}\left(2\rho_1\sqrt{\alpha\beta P_s P_r} + 2\rho_2\left(1 - e^R\right)\sqrt{\bar{\alpha}\bar{\beta}P_s P_r}\right) > 0$$

$$u^2 P_r\left(1 - \bar{\beta}e^R\right) + uk - \left(e^R\left(1 + a\bar{\alpha}P_s\right) - 1 - aP_s\right) < 0$$

and $f(u) < -1$ implies $u^2 P_r\left(1 - \bar{\beta}e^R\right) - uk - \left(e^R\left(1 + a\bar{\alpha}P_s\right) - 1 - aP_s\right) > 0$. Starting with $a < \frac{e^R - 1}{P_s\left(1 - \bar{\alpha}e^R\right)}$, $f(u) > 1$ holds for $0 < u < u_1$, where

$$u_1 = \frac{-k + \sqrt{k^2 + 4P_r\left(1-\bar{\beta}e^R\right)\left(e^R\left(1+a\bar{\alpha}P_s\right)-1-aP_s\right)}}{2P_r\left(1-\bar{\beta}e^R\right)}$$

and $f(u) < -1$ holds for $u > u_2$ where

$$u_2 = \frac{k + \sqrt{k^2 + 4P_r\left(1-\bar{\beta}e^R\right)\left(e^R\left(1+a\bar{\alpha}P_s\right)-1-aP_s\right)}}{2P_r\left(1-\bar{\beta}e^R\right)}.$$

Next, denote $a'$ as the positive solution to $k^2 = -4P_r\left(1-\bar{\beta}e^R\right)\left(e^R\left(1+a\bar{\alpha}P_s\right)-1-aP_s\right)$ which results in determinant being equal to zero. Since $k^2(a)$ is linear and positive while

$$-4P_r\left(1-\bar{\beta}e^R\right)\left(e^R-1\right)-aP_s\left(e^R\bar{\alpha}-1\right)4P_r\left(1-\bar{\beta}e^R\right)$$

increases linearly with $a$ and is negative at $a = 0$ assuming $\beta > \alpha$, then the solution exists if

$$\left(1-\bar{\alpha}e^R\right)\left(1-\bar{\beta}e^R\right) > \rho_1^2\alpha\beta + \rho_2^2\bar{\alpha}\bar{\beta}\left(1-e^R\right)^2 + 2\rho_1\rho_2\sqrt{\alpha\beta\bar{\alpha}\bar{\beta}}\left(1-e^R\right)$$

otherwise $a' = \infty$ as the determinant is positive for $a \geq 0$. Now, with $\frac{e^R-1}{P_s\left(1-\bar{\alpha}e^R\right)} < a < a'$, $f(u) \leq 1$ for all $u > 0$, however, $f(u) < -1$ iff $0 < u < u_3 \vee u > u_2$ where

$$u_3 = \frac{k - \sqrt{k^2 + 4P_r\left(1-\bar{\beta}e^R\right)\left(e^R\left(1+a\bar{\alpha}P_s\right)-1-aP_s\right)}}{2P_r\left(1-\bar{\beta}e^R\right)}.$$

Finally, for $a > a'$, $f(u) < -1, \forall u \geq 0$. It follows that

$$\int_0^\infty 2ue^{-u^2} P(\cos(\phi) > f(u)) du = \begin{cases} \int_{u_1}^{u_2} 2ue^{-u^2}\left(\frac{1}{2} - \frac{1}{\pi}\arcsin(f(u))\right) du + e^{-u_2^2}, a < \frac{e^R-1}{P_s\left(1-\bar{\alpha}e^R\right)} \\ \int_{u_3}^{u_2} 2ue^{-u^2}\left(\frac{1}{2} - \frac{1}{\pi}\arcsin(f(u))\right) du + 1 - e^{-u_3^2} + e^{-u_2^2}, \frac{e^R-1}{P_s\left(1-\bar{\alpha}e^R\right)} < a < a' \\ 1, a > a' \end{cases} \quad (16).$$

The evaluation of the expression for $k < 0$ follows along similar lines. If $a < \frac{e^R-1}{P_s\left(1-\bar{\alpha}e^R\right)}$ then $f(u) > 1$ holds for $u > u_1$ and $f(u) < -1$ holds for $0 < u < u_2$, $u_1, u_2$ are defined as for $k > 0$ case. If $\frac{e^R-1}{P_s\left(1-\bar{\alpha}e^R\right)} < a < a'$, then $f(u) > 1$ for $0 < u < u_3 \vee u > u_1$ where

$$u_3 = \frac{-k - \sqrt{k^2 + 4P_r\left(1-\bar{\beta}e^R\right)\left(e^R\left(1+a\bar{\alpha}P_s\right)-1-aP_s\right)}}{2P_r\left(1-\bar{\beta}e^R\right)}$$

and $f(u) > -1, \forall u \geq 0$. Finally, for $a > a'$, $f(u) > 1, \forall u \geq 0$. It follows that for this case

$$\int_0^\infty 2ue^{-u^2} P(\cos(\phi) > f(u)) du = \begin{cases} \int_{u_2}^{u_1} 2ue^{-u^2}\left(\frac{1}{2}+\frac{1}{\pi}a\sin(f(u))\right)du + e^{-u_1^2}, a < \frac{e^R-1}{P_s(1-\bar{\alpha}e^R)} \\ \int_{u_3}^{u_1} 2ue^{-u^2}\left(\frac{1}{2}+\frac{1}{\pi}a\sin(f(u))\right)du + 1 - e^{-u_3^2} + e^{-u_1^2}, \frac{e^R-1}{P_s(1-\bar{\alpha}e^R)} < a < a' \\ 1, a > a' \end{cases} \quad (17).$$

Summarizing, for $k > 0$, (13) can be expressed as

$$P(I(x_{s,1}, x_{r,1} : y_d) > R) = \int_0^\infty \left(\int_0^\infty 2ue^{-u^2} P(\cos(\phi) > f(u)) du\right) e^{-a} da$$

$$= \int_0^{\frac{e^R-1}{P_s(1-\bar{\alpha}e^R)}} \left(\frac{1}{2}\left(e^{-u_1^2} + e^{-u_2^2}\right) - \int_{u_1}^{u_2} \frac{2}{\pi} ue^{-u^2} \arcsin(f(u)) du\right) e^{-a} da + \int_{\frac{e^R-1}{P_s(1-\bar{\alpha}e^R)}}^{a'} \left(1 - \frac{1}{2}\left(e^{-u_3^2} - e^{-u_2^2}\right) - \int_{u_3}^{u_2} \frac{2}{\pi} ue^{-u^2} \arcsin(f(u)) du\right) e^{-a} da + e^{-a'}$$

(18)

and the substitution $u_1 \Leftrightarrow u_2$ gives the expression for $k < 0$.

For the second term of (42) the solution is similar to the one derived for the single layer cut-set bound [2], so we only state it here with the appropriate adaptations. Define using the previous definitions for $a, u$

$$g(u) \triangleq \frac{e^{R_2} - 1 - a\bar{\alpha}P_s - u^2\bar{\beta}P_r}{2\rho_2\sqrt{a\bar{\alpha}P_s\bar{\beta}P_r}u}, t \triangleq \sqrt{\frac{e^{R_2} - 1 - a(1-\rho_2^2)\bar{\alpha}P_s}{\bar{\beta}P_r}}, n \triangleq \rho_2\sqrt{\frac{a\bar{\alpha}P_s}{\bar{\beta}P_r}}$$

and denote $u_1 = -n+t, u_2 = n+t, u_3 = n-t$, then

$$P(I(x_{s,2}, x_{r,2} : y_d) > R) = \int_0^\infty \left(\int_0^\infty 2ue^{-u^2} P(\cos(\phi) > g(u)) du\right) e^{-a} da = \int_0^{\frac{e^R-1}{\bar{\alpha}P_s}} \left(\frac{1}{2}\left(e^{-u_1^2} + e^{-u_2^2}\right) - \int_{u_1}^{u_2} \frac{2}{\pi} ue^{-u^2} a\sin(g(u)) du\right) e^{-a} da$$

$$+ \int_{\frac{e^R-1}{\bar{\alpha}P_s}}^{\frac{e^R-1}{(1-\rho_2^2)\bar{\alpha}P_s}} \left(1 - \frac{1}{2}\left(e^{-u_3^2} - e^{-u_2^2}\right) - \int_{u_3}^{u_2} \frac{2}{\pi} ue^{-u^2} a\sin(g(u)) du\right) e^{-a} da + e^{-\frac{e^R-1}{(1-\rho_2^2)\bar{\alpha}P_s}}$$

(19).

By employing (18) we can compute explicitly the probability of decoding $R_1$, but the intersection term $P(R_1^P \cap R_2^P)$ required for the decoding probability of $R_2$ does not lend itself analytically tractable. Expression (19) can be used as an upper bound for the decoding probability of $R_2$ as

$$R_{av} = R_1 \cdot P(R_1^P) + R_2 \cdot P(R_1^P \cap R_2^P), P(R_1^P \cap R_2^P) \leq P(R_2^P).$$

Due to the complexity of the expressions, we turn to the Monte-Carlo technique in order to evaluate the achievable throughput for a given SISO attempted rates. In Fig. 7 displayed are the achievable rates for all correlation coefficients under the oblivious setting. Several important features can be witnessed. High $\rho_1$ are non-feasible which is in agreement with (11) being a decreasing function of the parameter. The exact behavior is dictated by the location of the vertices of the hyperbola, but in general a higher $\rho_2$ will result in higher non-feasible $\rho_1$. The smallest non-feasible $\rho_1$ is given by $\rho^* = \sqrt{\frac{1+P_sQ - e^{R_1}(1+Q\bar{\alpha}P_s)}{P_sQ\alpha\beta}}$ for $F > 0$.

Obviously, if $\rho^* > 1$ then every $\rho_1$ is feasible. The limitation on $\rho_2$ mostly comes from the second layer. It describes the limit from which on the second layer is non-decodable and therefore a major drop in the achievable throughput is expected. From the simulations, the decoding probability of the first layer also decreases for $\rho_2 \to 1$, so we are left with a very low throughput. The maximal throughput for all cases checked is attained with $\rho_1 = \rho_2 = 0$, but in the vicinity of the origin the rate is almost independent of the correlation coefficients. This is probably due to the use of an oblivious rate which sets low coefficients to terms involving $\rho_1, \rho_2$.

Note that in both cases the maximum is located on the boundary of the feasible coefficients region. We may draw some reasoning of this by examination of the MISO term in (39). Fixing $\rho_1$, if the resulting function is increasing with $\rho_2$, then the maximum for each fixed $\rho_1$ is attained for the highest feasible $\rho_2$, i.e on the boundary. If the function is decreasing with $\rho_2$, the maximum is attained for the lowest feasible $\rho_2$ this time, being on the boundary as well. Similar claim can be made by swapping the roles of $\rho_1, \rho_2$. In order to show the properties discussed, we need to examine $\frac{\partial P(I(x_s, x_r : y_d) > R)}{\partial \rho_1}$ and $\frac{\partial P(I(x_s, x_r : y_d) > R)}{\partial \rho_2}$ and verify the monotony w.r.t to the parameters. Strong numerical evidence in favor of the mutual information being monotonic stems from the numerous cases checked in this work.

## V. CONCLUSION

In this paper, part II of a two-part series dealing with the incorporation of broadcast approach into oblivious cooperation protocols for the relay channel, we addressed the Block-Markov decode-and-forward scheme. For the single layer transmission, we have shown that MISO rates are achievable for oblivious cooperation conditioned on a positive (in dB) collocation gain between the source and the relay, thus largely improving upon the SDF requiring infinite gains for this. In addition, we have established a conjecture regarding the optimality of the uncorrelated transmission between the source and the relay in terms of throughput and a theorem regarding the correlation uncertainty region, extending [2].

For the two-layered transmission, we have formulated the general rates attainable by the scheme and shown that the MISO rates are feasible here as well for the oblivious cooperation, leading to even more pronounced gains comparing to the SDF which was interference limited in this case. In fact, this applies to any number of discrete layers used. With correlated transmission applied for $N = 2$, analysis of the general rate expression was carried out and the probability of decoding the first layer was explicitly computed analytically, and an upper bound to the overall decoding probability was provided.

Further research of the subject will include determining the optimal correlation coefficients for the multi-layer transmission, extension of the achievable rates expressions to include the variable power allocation for more than two layers and the continuous broadcasting upper bounds. Truly oblivious protocols such as the compress-and-forward [2],[10] in a broadcasting framework are of major interest as well.

## APPENDIX A

### CONJECTURE 1

In order to show that the optimal throughput is attained for $\rho = 0$ we use the results by Katz and Shamai [4] for the successful decoding probability of a 2x1 MISO with correlation constraint $\rho_{max}$ and two assumptions regarding the throughput function :

1. The throughput function with $\rho = 0$ is uni-modal with a global maximum. This is proven further on.
2. The throughput is continuous at the point $R_0 = \log(1+P), P \triangleq P_s + P_r$. Note that this is the boundary point between the regions for which $\rho^{opt} = 0, R < R_0$ and $\rho^{opt} = 0$ or $\rho^{opt} = \rho_{max}, R > R_0$.

Assuming $R_0$ is decodable as otherwise the proof is trivial, we start with the case $P_s = P_r$ and compute the derivative of the throughput function which is $R \cdot \bar{P}_{suc}^{IP}(R, P_s, P_r, \rho_{max})$ by definition. From [4], for $R_0 = \log(1+P)$,

$$\bar{P}_{suc}^{IP}(R, P_s, P_r, \rho_{max}) = r(P_s, P_r, e^R - 1), r(x, y, c) = \begin{cases} \dfrac{xe^{-\frac{c}{x}} - ye^{-\frac{c}{y}}}{x - y}, x \neq y \\ \left(1 + \dfrac{c}{x}\right)e^{-\frac{c}{x}}, x = y \end{cases}.$$

so the first derivative of the throughput is computed as

$$\frac{\partial f_{TP}}{\partial R} = \frac{\partial}{\partial R}\left(R \cdot r(P_s, P_s, e^R - 1)\right) = r(x, x, c) - \frac{cRe^R}{x^2}e^{-\frac{c}{x}} = \left(1 + \frac{c}{x}\right)e^{-\frac{c}{x}} - \frac{cRe^R}{x^2}e^{-\frac{c}{x}}. \quad (20)$$

A necessary condition to the sub-optimality of $R = R_0$ is $f_{TP}(R_0)' < 0$ since this coupled with the first assumption would establish a lower rate as the optimal. By definition of the involved quantities, the inequality becomes

$$\frac{3P_s}{2} < \log(1+2P_s)(1+2P_s). \quad (21)$$

To prove (21), we use the inequality $\dfrac{x}{1+x} < \log(1+x), x > -1$ with the definition $x = 2P_s$, therefore the derivative is negative making the optimal rate satisfy $R < R_0$. But, from [4], the optimal correlation for $R < R_0$ is $\rho^{opt} = 0$.

Now, we turn to the general case of $P_s \neq P_r$. Similarly, we compute the derivative as

$$\frac{\partial f_{TP}}{\partial R} = \frac{xe^{-\frac{c}{x}} - ye^{-\frac{c}{y}}}{x-y} + R \cdot e^R \left( \frac{e^{-\frac{c}{y}} - e^{-\frac{c}{x}}}{x-y} \right) \quad (22)$$

and once again we need to prove the function is decreasing at $R = R_0$. Rearranging, we get

$$\frac{P_s e^{-\frac{P_r}{P_s}} - P_r e^{-\frac{P_s}{P_r}}}{e^{-\frac{P_r}{P_s}} - e^{-\frac{P_s}{P_r}}} < \log(1 + P_s + P_r)(1 + P_s + P_r). \quad (23)$$

Suppose first that $P_r > P_s$, then $\frac{P_s e^{-\frac{P_r}{P_s}} - P_r e^{-\frac{P_s}{P_r}}}{e^{-\frac{P_r}{P_s}} - e^{-\frac{P_s}{P_r}}} < P_r < (P_s + P_r) < \log(1 + P_s + P_r)(1 + P_s + P_r)$ by the inequality previously used where this time $x = P_s + P_r$ and the derivative is negative. Finally, suppose $P_s > P_r$. We claim that in this case the following inequality takes place

$$\frac{P_s e^{-\frac{P_r}{P_s}} - P_r e^{-\frac{P_s}{P_r}}}{e^{-\frac{P_r}{P_s}} - e^{-\frac{P_s}{P_r}}} < (P_s + P_r). \quad (24)$$

To prove this, it suffices to show that $P_r e^{-\frac{P_r}{P_s}} - P_s e^{-\frac{P_s}{P_r}} > 0$ by arranging the terms of the previous form. Define $n \triangleq 0 \leq \frac{P_r}{P_s} < 1$, then (24) is equivalent to $n \cdot e^{-n} \geq e^{-\frac{1}{n}}$. Consider now the function $f(n) = \frac{e^{-\frac{1}{n}}}{n}$ and its derivatives

$$f'(n) = \frac{(1-n)}{n^3} e^{-\frac{1}{n}}, f''(n) = \left( \frac{2}{n^3} - \frac{4}{n^4} + \frac{1}{n^5} \right) e^{-\frac{1}{n}}. \quad (25)$$

Referring to (25), the function attains an extremum at $n=1$ which is a maximum with a value of $e^{-1}$, thus $ne^{-1} > e^{-\frac{1}{n}}$, but $e^{-n} > e^{-1}, n < 1$ completing the proof.

Next, we prove assumption 1 by a direct examination of the throughput function for $\rho = 0$. Starting with the simpler case of $P_s = P_r$, the throughput and its derivative are computed as

$$f_{TP}(R) = R \cdot \left( 1 + \frac{c}{P_s} \right) e^{-\frac{c}{P_s}}, c \triangleq e^R - 1, g(R) = f'_{TP}(R) = \left( 1 + \frac{c}{P_s} - \frac{cR \cdot e^R}{P_s^2} \right) e^{-\frac{c}{P_s}}. \quad (26)$$

An extremum of the throughput is attained for $g(R) = 0$. Solving, we get

$$\left( 1 + \frac{e^R - 1}{P_s} \right) = R \cdot e^R \frac{(e^R - 1)}{P_s^2}, m(R) = P_s^2 + P_s(e^R - 1) = R \cdot e^R (e^R - 1) = n(R). \quad (27)$$

It is easily verified that both $m(R), n(R)$ are monotonically increasing and positive functions satisfying $m(0) > n(0), m(R)_{R \to \infty} < n(R)_{R \to \infty}$. Furthermore,

$$m''(R) = P_s e^R > 0, n''(R) = 4(1+R)e^{2R} - 2(1+R)e^R > 0 \quad (28)$$

and therefore both functions are strictly convex $\cup$. It is easily verified that there exists an intersection point for which $m(R) = n(R)$. We need to show that there is a unique intersection point as in general there can be an odd number of intersections $2k+1, k \geq 0$. To prove this, for any fixed $R$ we look for the $P_s$ which will satisfy (27). This unique $P_s$ is obtained by solving a quadratic equation to get

$$P_s(R) = \frac{-(e^R - 1) + \sqrt{(e^R - 1)^2 + 4R(e^R - 1)e^R}}{2}.$$

The function $P_s(R)$ is strictly increasing since

$$P_s'(R) = \frac{e^R}{2}\left(\frac{6(e^R - 1) + 4R(e^R - 1) + 4e^{2R}R}{2\sqrt{(e^R - 1)^2 + 4R(e^R - 1)e^R}} - 1\right) > 0, R > 0, P_s'(R)_{R \to 0^+} = \frac{\sqrt{5} - 1}{2} > 0$$

and is therefore invertible, that is we can compute the unique $R$ for each $P_s$ which is the solution to (27) by simply reflecting the function across the line $P_s = R$. This means that the function $f_{TP}(R)$ has only a single maximum with $\rho = 0$.

Suppose now $P_s \neq P_r$ and the throughput function has the form of

$$f_{TP}(R) = R \cdot \left(\frac{P_s e^{-\frac{c}{P_s}} - P_r e^{-\frac{c}{P_r}}}{P_s - P_r}\right), c \triangleq e^R - 1. \quad (29)$$

Computing the derivative, we get (22) and by equating the derivative to zero

$$\left(\frac{P_s e^{-\frac{c}{P_s}} - P_r e^{-\frac{c}{P_r}}}{P_s - P_r}\right) = R \cdot e^R\left(\frac{e^{-\frac{c}{P_s}} - e^{-\frac{c}{P_r}}}{P_s - P_r}\right), m(R) = \left(\frac{P_s e^{-\frac{c}{P_s}} - P_r e^{-\frac{c}{P_r}}}{e^{-\frac{c}{P_s}} - e^{-\frac{c}{P_r}}}\right), R \cdot e^R = n(R). \quad (30)$$

Once again, the functions satisfy $m(R)_{R \to 0} \to \infty > n(R)_{R \to 0} = 0$ and therefore an intersection point must exist. The function $n(R)$ is strictly convex $\cup$ and monotonically increasing while

$$m'(R) = -\frac{\left(e^R e^{-\frac{c}{P_s}} e^{-\frac{c}{P_r}}(P_s - P_r)^2\right)}{P_s P_r \left(e^{-\frac{c}{P_s}} - e^{-\frac{c}{P_r}}\right)^2} < 0, m^*(R) = e^R e^{-\frac{c}{P_s}} e^{-\frac{c}{P_r}}(P_s - P_r)^2 \frac{\left(e^{-\frac{c}{P_r}}\left(e^R(P_s - P_r) + P_s P_r\right) + e^{-\frac{c}{P_s}}\left(e^R(P_s - P_r) - P_s P_r\right)\right)}{P_s^2 P_s^2 \left(e^{-\frac{c}{P_s}} - e^{-\frac{c}{P_r}}\right)^3} > 0$$

, so the intersection point must be unique.

So far, we have proven that for $\rho = 0$ the throughput function is uni-modal and attains its maximum for $R < R_0 = \log(1 + P)$. Let us now examine the throughput function for $\rho^{opt} = \rho_{\max}$ that can be stated as

$$f_{TP}(R) = R \cdot \left(\frac{P_0 e^{-\frac{c}{P_0}} - \overline{P}_0 e^{-\frac{c}{\overline{P}_0}}}{P_0 - \overline{P}_0}\right) = R \cdot P_d(R), P_0 \triangleq \frac{1}{2}\left[P + \sqrt{P^2 - 4\frac{c}{Q}P_r}\right], \overline{P}_0 = P - P_0 \quad (31)$$

While it is possible to explicitly compute the derivative of the throughput function under this condition despite the complexity, proving $f_{TP}(R_0) > f_{TP}(R' > R_0)$ is sufficient as this will ensure the maximum under $\rho_{max}$ is obtained for $R \leq R_0$, therefore being smaller than that of $\rho = 0$, and this is a less stringent requirement than uni-modality. Starting with $P_s = P_r$, define $R' \triangleq \log(1+\alpha P), 1 < \alpha \leq \frac{Q}{2}$ by which

$$c = \alpha P, P_0 = P_s\left(1+\sqrt{1-\frac{2\alpha}{Q}}\right), \overline{P_0} = P_s\left(1-\sqrt{1-\frac{2\alpha}{Q}}\right), P_0 - \overline{P_0} = 2P_s\sqrt{1-\frac{2\alpha}{Q}}.$$

and $Q > 2$ by $R_0$ achievability. Plugging $R_0, R'$ together with the resulting quantities into (31) leads to the following inequality

$$\frac{\log(1+\alpha P)}{\log(1+P)} \leq \left[\frac{\sqrt{1-\frac{2\alpha}{Q}}\left(\left(1+\sqrt{1-\frac{2}{Q}}\right)e^{-\frac{2}{1+\sqrt{1-\frac{2}{Q}}}} - \left(1-\sqrt{1-\frac{2}{Q}}\right)e^{-\frac{2}{1-\sqrt{1-\frac{2}{Q}}}}\right)}{\sqrt{1-\frac{2}{Q}}\left(\left(1+\sqrt{1-\frac{2\alpha}{Q}}\right)e^{-\frac{2\alpha}{1+\sqrt{1-\frac{2\alpha}{Q}}}} - \left(1-\sqrt{1-\frac{2\alpha}{Q}}\right)e^{-\frac{2\alpha}{1-\sqrt{1-\frac{2\alpha}{Q}}}}\right)}\right], 1 < \alpha \leq \frac{Q}{2} \quad (32)$$

For fixed $\alpha$ and $Q \to \infty$ the inequality hold as the RHS becomes $e^{\alpha-1} > \alpha$ and $\log(1+\alpha P) < \alpha \log(1+P)$ since $(1+x)^n > 1+nx, x > 0$. For $\alpha = 1$ an equality holds, hence for very low collocation gains we expect the difference to be marginal. In order to get some insight into (32), note that the RHS can be written as

$$\frac{k(1)}{k(\alpha)}, k(\alpha) = \frac{\left(1+\sqrt{1-\frac{2\alpha}{Q}}\right)e^{-\frac{2\alpha}{1+\sqrt{1-\frac{2\alpha}{Q}}}} - \left(1-\sqrt{1-\frac{2\alpha}{Q}}\right)e^{-\frac{2\alpha}{1-\sqrt{1-\frac{2\alpha}{Q}}}}}{2\sqrt{1-\frac{2\alpha}{Q}}}, 1 < \alpha \leq \frac{Q}{2}, k'(\alpha) < 0 \quad (33)$$

and therefore the maximum is attained for $\alpha = 1$ and the RHS of (32) $\geq 1$, with $k(\alpha)$ decreasing at an exponential rate of $\alpha \cdot r, -2 < r < -1$. Consider now $g(\alpha) = \alpha k(\alpha)$. Proving $g(\alpha)$ is a decreasing function will lead to the desired inequality as $g(\alpha) = \alpha k(\alpha) < g(1) = k(1)$. The sign of the derivative $g'(\alpha) = k(\alpha) + \alpha k'(\alpha)$ at an arbitrary point is hard to determine, but at the two boundary points $\alpha \to 1, \frac{Q}{2}$ one can verify that $g'(\alpha) < 0$, meaning that $g(1)$ is not a minimum over the relevant range. The complexity of the inequality requires a numerical verification for finite values of $\alpha, Q$ presented in Fig. 8. Since $\frac{k(1)}{k(\alpha)} > \alpha$, $\rho_{max}$ can not lead to the maximal throughput for $P_s = P_r$ as the maximum is obtained for $R < R_0$.

Finally, we examine the case $P_s \neq P_r$. Under those conditions, with $R_0 = \log(1+P)$ the power allocation expressions becomes

$$P_0 = \frac{1}{2}\left[P + \sqrt{P^2 - \frac{4PP_r}{Q}}\right], \overline{P_0} = \frac{1}{2}\left[P - \sqrt{P^2 - \frac{4PP_r}{Q}}\right], P_0 - \overline{P_0} = \sqrt{P^2 - \frac{4PP_r}{Q}}$$

and the throughput becomes

$$f^{TP}(R_0) = R_0 \left( \frac{\left(P + \sqrt{P^2 - \frac{4PP_r}{Q}}\right) e^{-\frac{2P}{\left(P + \sqrt{P^2 - \frac{4PP_r}{Q}}\right)}} - \left(P - \sqrt{P^2 - \frac{4PP_r}{Q}}\right) e^{-\frac{2P}{\left(P - \sqrt{P^2 - \frac{4PP_r}{Q}}\right)}}}{2\sqrt{P^2 - \frac{4PP_r}{Q}}} \right) \quad (34)$$

while for $R' = \log(1+\alpha P), 1 < \alpha < \frac{P_s Q}{P}$ the throughput is

$$f^{TP}(R') = \frac{R'\left(P + \sqrt{P^2 - \frac{4\alpha PP_r}{Q}}\right) e^{-\frac{2\alpha P}{\left(P + \sqrt{P^2 - \frac{4\alpha PP_r}{Q}}\right)}}}{2\sqrt{P^2 - \frac{4\alpha PP_r}{Q}}} - \frac{R'\left(P - \sqrt{P^2 - \frac{4\alpha PP_r}{Q}}\right) e^{-\frac{2\alpha P}{\left(P - \sqrt{P^2 - \frac{4\alpha PP_r}{Q}}\right)}}}{2\sqrt{P^2 - \frac{4\alpha PP_r}{Q}}}. \quad (35)$$

and we wish to show that $\frac{f_{TP}(R')}{f_{TP}(R_0)} < 1$ to establish the optimality of $\rho = 0$. For this, let us define $s \triangleq \frac{PQ}{2P_r} = \frac{(P_s + P_r)^2}{2P_s P_r} > 2$ due to the achievability of $R_0$, and rewrite the throughput inequality in an equivalent form of (36) where $1 < \alpha < \frac{P_s Q}{P} \leq \frac{s}{2} = \frac{PQ}{4P_r}$. But, (36) is exactly (32) with $s$ replacing $Q$ as a more general definition and $\alpha$ confined to a sub-interval relatively to the upper bound (for $P_s \neq P_r$ $\alpha < \frac{PQ}{P_s} \leq \frac{s}{2}$, for $P_s = P_r$ $\alpha \leq \frac{s|_{P_s = P_r} = Q}{2}$ ), completing the proof □.

$$\frac{\log(1+\alpha P)}{\log(1+P)} \leq \left[ \frac{\sqrt{1-\frac{2\alpha}{s}}\left(\left(1+\sqrt{1-\frac{2}{s}}\right)e^{-\frac{2}{\left(1+\sqrt{1-\frac{2}{s}}\right)}} - \left(1-\sqrt{1-\frac{2}{s}}\right)e^{-\frac{2}{\left(1-\sqrt{1-\frac{2}{s}}\right)}}\right)}{\sqrt{1-\frac{2}{s}}\left(\left(1+\sqrt{1-\frac{2\alpha}{s}}\right)e^{-\frac{2\alpha}{\left(1+\sqrt{1-\frac{2\alpha}{s}}\right)}} - \left(1-\sqrt{1-\frac{2\alpha}{s}}\right)e^{-\frac{2\alpha}{\left(1-\sqrt{1-\frac{2\alpha}{s}}\right)}}\right)} \right] \quad (36)$$

Summarizing, we have shown that the throughput function with $\rho = \rho_{\max}$ attains its maximum at $R < R_0$, meaning that the highest throughput is attained with $\rho = 0$ optimal for $R < R_0$.

APPENDIX B

THEOREM 2

For the first layer, the source-relay channel mutual information becomes

$$I(x_{s,1}:y_r|x_r) = H(y_r|x_r) - H(y_r|x_{s,1},x_r).$$

Solving the first term, we get

$$H(y_r|x_r) = H(y_r,x_r) - H(x_r) = H(\sqrt{Q}(x_{s,1}+x_{s,2})+z_r,x_r) - H(x_r)$$

$$= \log\left|\pi e \begin{pmatrix} 1+P_sQ & \sqrt{Q\alpha\beta P_s P_r}\rho_1^* + \sqrt{Q\bar{\alpha}\bar{\beta}P_s P_r}\rho_2^* \\ \sqrt{Q\alpha\beta P_s P_r}\rho_1 + \sqrt{Q\bar{\alpha}\bar{\beta}P_s P_r}\rho_2 & P_r \end{pmatrix}\right| - \log(\pi e P_r) = \log\left(\pi e\left(1+P_sQ\left(1-\alpha\beta|\rho_1|^2-\bar{\alpha}\bar{\beta}|\rho_2|^2-2\sqrt{\alpha\beta\bar{\alpha}\bar{\beta}}\Re_e(\rho_1\rho_2^*)\right)\right)\right)$$

, similarly for the second term

$$H(y_r|x_{s,1},x_r) = H(\sqrt{Q}x_{s,2}+z_r|x_r) = \log\left|\pi e \begin{pmatrix} 1+Q\bar{\alpha}P_s & \sqrt{Q\bar{\alpha}\bar{\beta}P_s P_r}\rho_2^* \\ \sqrt{Q\bar{\alpha}\bar{\beta}P_s P_r}\rho_2 & \bar{\beta}P_r \end{pmatrix}\right| - \log(\pi e \bar{\beta}P_r) = \log\left(\pi e\left(1+Q\bar{\alpha}P_s\cdot(1-|\rho_2|^2)\right)\right),$$

and overall for the first layer

$$I(x_{s,1}:y_r|x_r) = \log\left(\frac{\left(1+P_sQ\left(1-\alpha\beta|\rho_1|^2-\bar{\alpha}\bar{\beta}|\rho_2|^2-2\sqrt{\alpha\beta\bar{\alpha}\bar{\beta}}\Re_e(\rho_1\rho_2^*)\right)\right)}{1+Q\bar{\alpha}P_s\cdot(1-|\rho_2|^2)}\right) \quad (37)$$

For the MISO term, one has

$$I(x_{s,1},x_{r,1}:y_d) = H(y_d) - H(y_d|x_{s,1},x_{r,1}) = H(h_s(x_{s,1}+x_{s,2}) + h_r(x_{r,1}+x_{r,2}) + z_d) - H(h_s(x_{s,1}+x_{s,2}) + h_r(x_{r,1}+x_{r,2}) + z_d|x_{s,1},x_{r,1})$$

$$H(h_s(x_{s,1}+x_{s,2}) + h_r(x_{r,1}+x_{r,2}) + z_d) = \log\left(\pi e\left(1+v_sP_s+v_rP_r+2\left(\sqrt{\alpha\beta P_s P_r}\Re_e(\rho_1 h_s h_r^*)+\sqrt{\bar{\alpha}\bar{\beta}P_s P_r}\Re_e(\rho_2 h_s h_r^*)\right)\right)\right)$$

$$H(h_s(x_{s,1}+x_{s,2}) + h_r(x_{r,1}+x_{r,2}) + z_d|x_{s,1},x_{r,1}) = \log\left(\pi e\left(1+v_s\bar{\alpha}P_s+v_r\bar{\beta}P_r+2\sqrt{\bar{\alpha}\bar{\beta}P_s P_r}\Re_e(\rho_2 h_s h_r^*)\right)\right)$$

and therefore

$$I(x_{s,1},x_{r,1}:y_d) = \log\left(\frac{1+v_sP_s+v_rP_r+2\left(\sqrt{\alpha\beta P_s P_r}\Re_e(\rho_1 h_s h_r^*)+\sqrt{\bar{\alpha}\bar{\beta}P_s P_r}\Re_e(\rho_2 h_s h_r^*)\right)}{1+v_s\bar{\alpha}P_s+v_r\bar{\beta}P_r+2\sqrt{\bar{\alpha}\bar{\beta}P_s P_r}\Re_e(\rho_2 h_s h_r^*)}\right) \quad (38).$$

From (6), the decoding probability of the first layer is given by

$$P\left(\left(\log\left(\frac{1+P_sQ\left(1-\alpha\beta|\rho_1|^2-\bar{\alpha}\bar{\beta}|\rho_2|^2-2\sqrt{\alpha\beta\bar{\alpha}\bar{\beta}}\Re_e(\rho_1\rho_2^*)\right)}{1+Q\bar{\alpha}P_s\cdot(1-|\rho_2|^2)}\right) > R_1\right) \cap \left(\log\left(\frac{1+v_sP_s+v_rP_r+2\left(\sqrt{\alpha\beta P_s P_r}\Re_e(\rho_1 h_s h_r^*)+\sqrt{\bar{\alpha}\bar{\beta}P_s P_r}\Re_e(\rho_2 h_s h_r^*)\right)}{1+v_s\bar{\alpha}P_s+v_r\bar{\beta}P_r+2\sqrt{\bar{\alpha}\bar{\beta}P_s P_r}\Re_e(\rho_2 h_s h_r^*)}\right) > R_1\right)\right) \quad (39)$$

For the second layer, assuming that the first layer has been successfully decoded, one has

$$I(x_{s,2}:y_r|x_r) = H(y_r|x_r) - H(y_r|x_{s,2},x_r), H(y_r|x_r) = H(y_r,x_r) - H(x_r) = H(\sqrt{Q}(x_{s,2})+z_r,x_r) - H(x_r)$$

$$= \log\left|\pi e \begin{pmatrix} 1+Q\bar{\alpha}P_s & \sqrt{Q\bar{\alpha}\bar{\beta}P_s P_r}\rho_2^* \\ \sqrt{Q\bar{\alpha}\bar{\beta}P_s P_r}\rho_2 & \bar{\beta}P_r \end{pmatrix}\right| - \log(\pi e \bar{\beta}P_r) = \log\left(\pi e\left(1+Q\bar{\alpha}P_s(1-|\rho_2|^2)\right)\right),$$

$$H(y_r|x_{s,2},x_r) = \log(\pi e)$$

leading to

$$I(x_{s,2}:y_r|x_r) = \log\left(1+Q\bar{\alpha}P_s(1-|\rho_2|^2)\right). \quad (40)$$

Finally,

$$I(x_{s,2}, x_{r,2} : y_d) = H(y_d) - H(y_d | x_{s,2}, x_{r,2}) = H(h_s(x_{s,2}) + h_r(x_{r,2}) + z_d) - H(h_s(x_{s,2}) + h_r(x_{r,2}) + z_d | x_{s,2}, x_{r,2})$$

$$H(h_s(x_{s,2}) + h_r(x_{r,2}) + z_d) = \log\left(\pi e\left(1 + v_s \bar{\alpha} P_s + v_r \bar{\beta} P_r + 2\sqrt{\bar{\alpha}\bar{\beta} P_s P_r} \Re_e(\rho_2 h_s h_r^*)\right)\right)$$

$$H(h_s(x_{s,2}) + h_r(x_{r,2}) + z_d | x_{s,2}, x_{r,2}) = \log(\pi e)$$

and therefore

$$I(x_{s,2}, x_{r,2} : y_d) = \log\left(1 + v_s \bar{\alpha} P_s + v_r \bar{\beta} P_r + 2\sqrt{\bar{\alpha}\bar{\beta} P_s P_r} \Re_e(\rho_2 h_s h_r^*)\right) \quad (41)$$

The decoding probability of the second layer conditioned on successful cancellation of the first appears as

$$P\left(\left(\log\left(1 + Q\bar{\alpha} P_s \left(1 - |\rho_2|^2\right)\right) > R_2\right) \cap \left(\log\left(1 + v_s \bar{\alpha} P_s + v_r \bar{\beta} P_r + 2\sqrt{\bar{\alpha}\bar{\beta} P_s P_r} \Re_e(\rho_2 h_s h_r^*)\right) > R_2\right)\right) \quad (42).$$

completing the proof. □

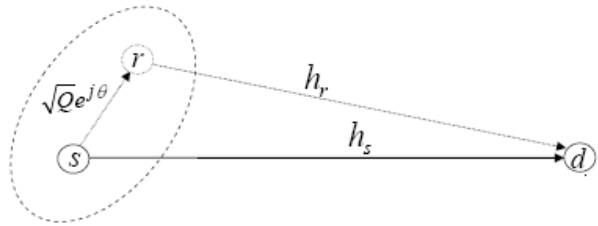

Fig. 1. A source-relay collocated network

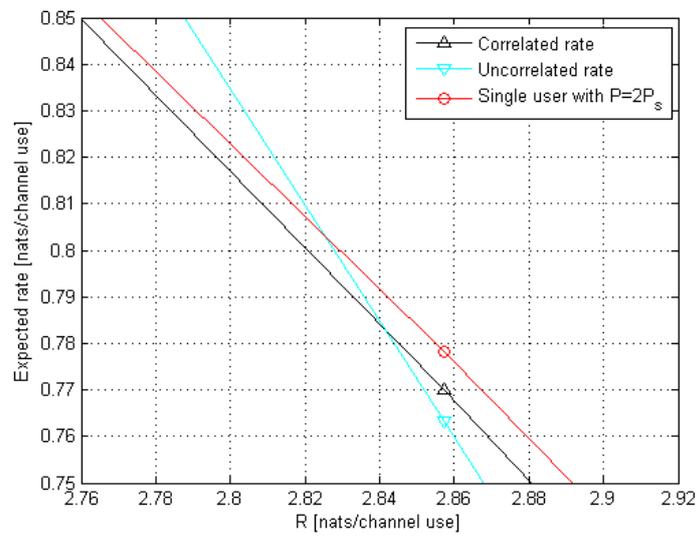

Fig. 2. Demonstration of Thm 1. for finite collocation gain of $Q = 10$dB, $P_s = P_r = 8$dB. The uncertainty region for the optimal correlation is reduced to the vicinity of $R_0$ when $P_s = P_r$ and vanishes when $Q \to \infty$.

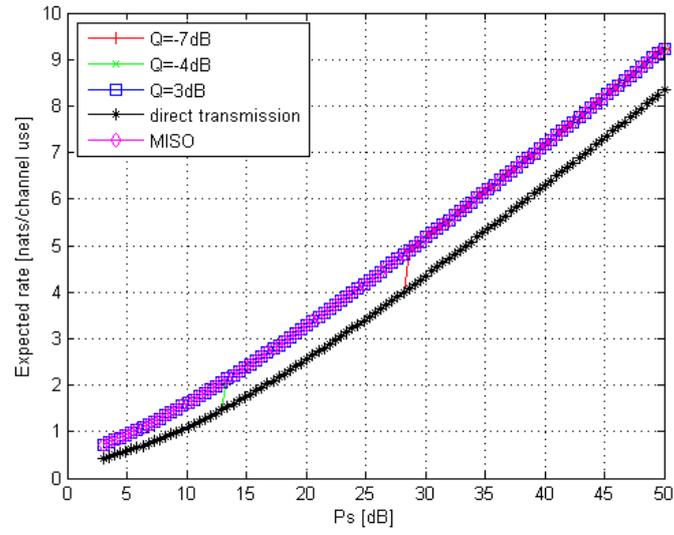

Fig. 3. Single layer oblivious BM rates as a function of source power $P_s$, relay power $P_r = P_s$ and the collocation gain $Q$.

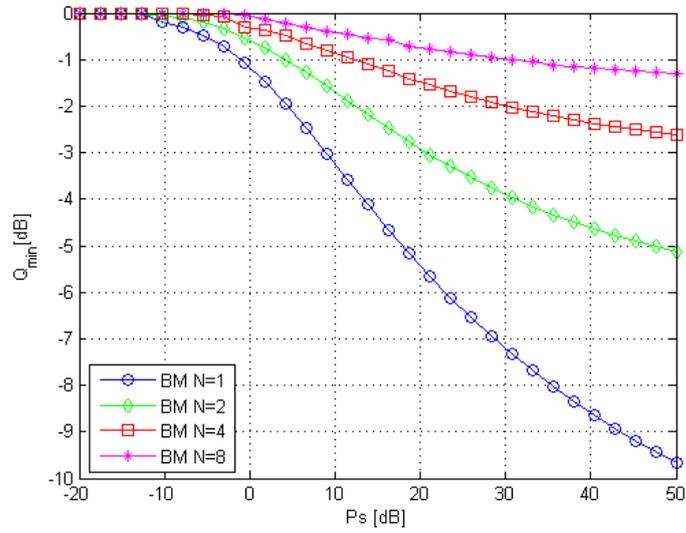

Fig. 4. Minimal $Q$ for MISO rate achievability for $N = 1, 2, 4, 8$ layers from (5), (10).

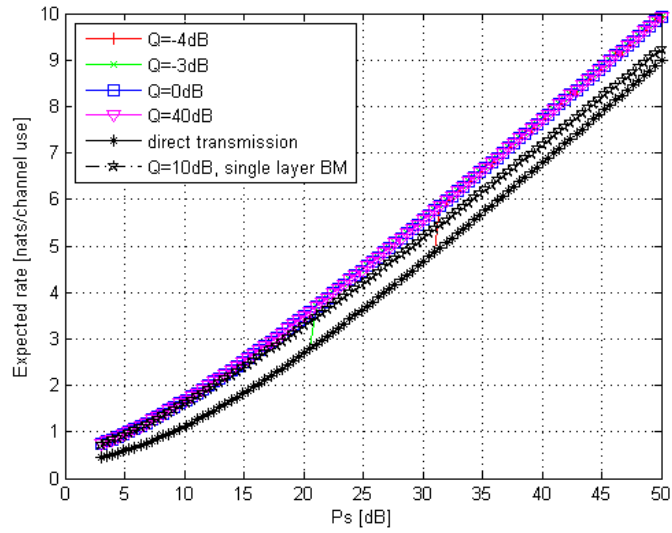

Fig. 5. Two layer oblivious uncorrelated BM achievable rates under unequal antenna layering power distribution as a function of source power $P_s$, relay power $P_r = P_s$ and the collocation gain $Q$.

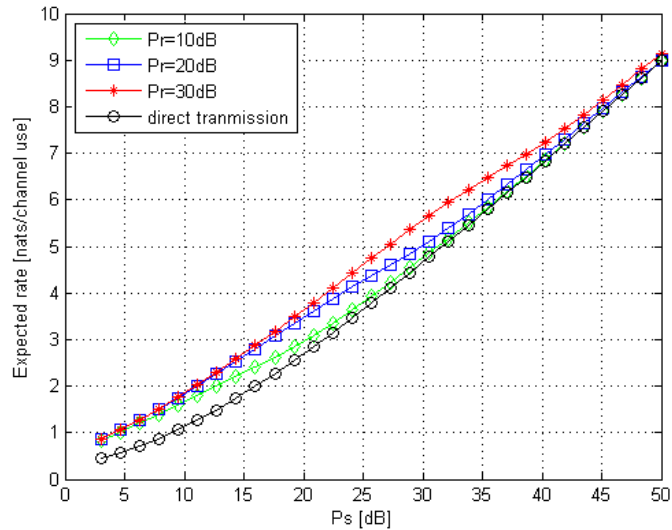

Fig. 6 . Two layer oblivious uncorrelated BM achievable rates under unequal antenna layering power distribution as a function of source power $P_s$, fixed relay power $P_r$ and the collocation gain $Q = 10$dB.

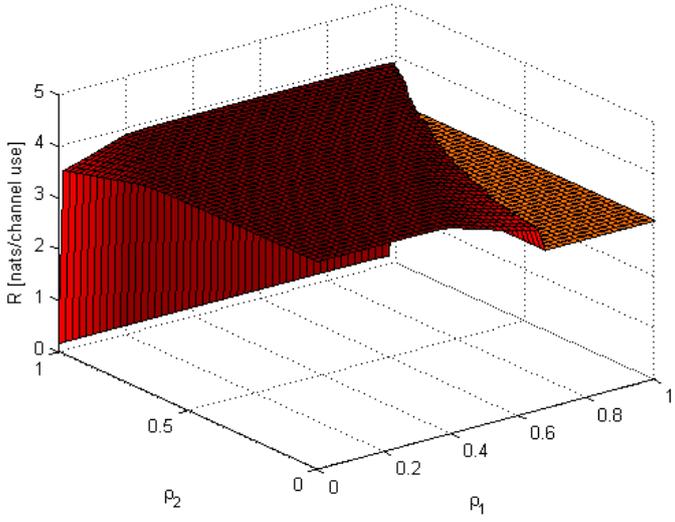

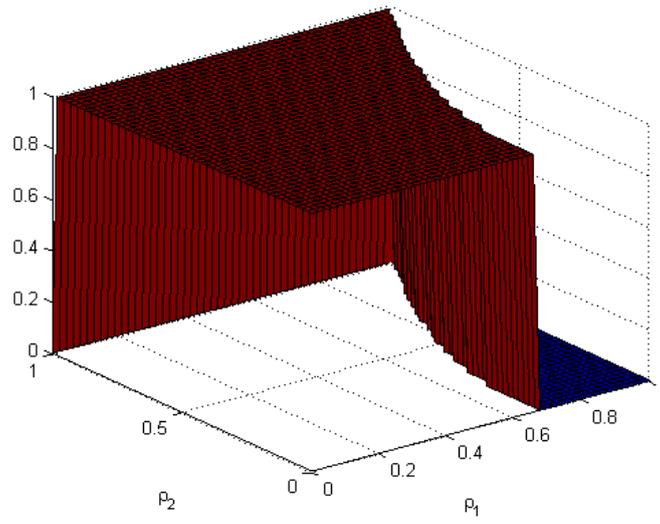

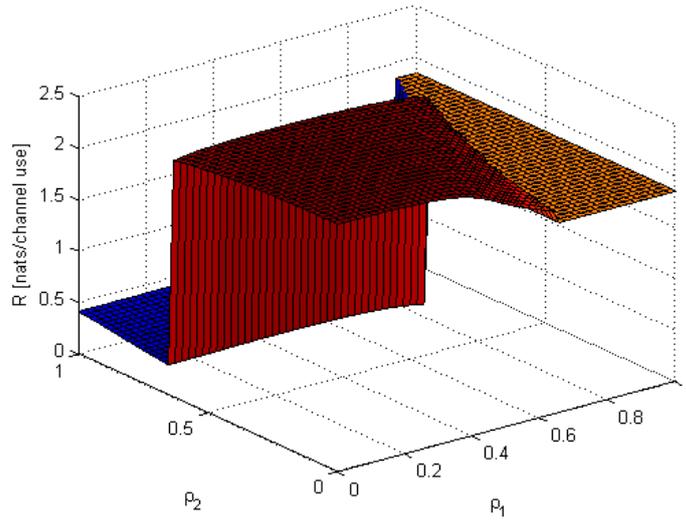

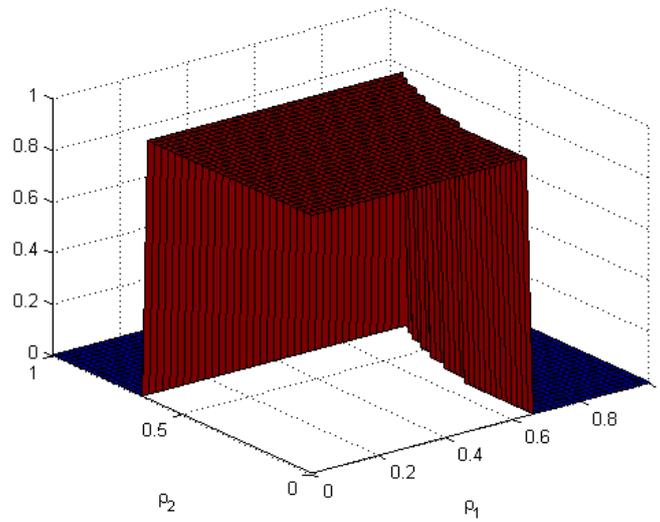

Fig. 7. Two layer oblivous BM achievable rates under equal antenna layering power distirbution for various correlation coefficients and the feasible correlation coefficients area indicator function $\mathbf{1}(\rho_1,\rho_2)$ from (11) (a,b) $P_s = 22$dB , $P_r = 30$dB, $Q = 40$dB, (c,d) $P_s = 15$dB , $P_r = 12$dB, $Q = 0$dB.

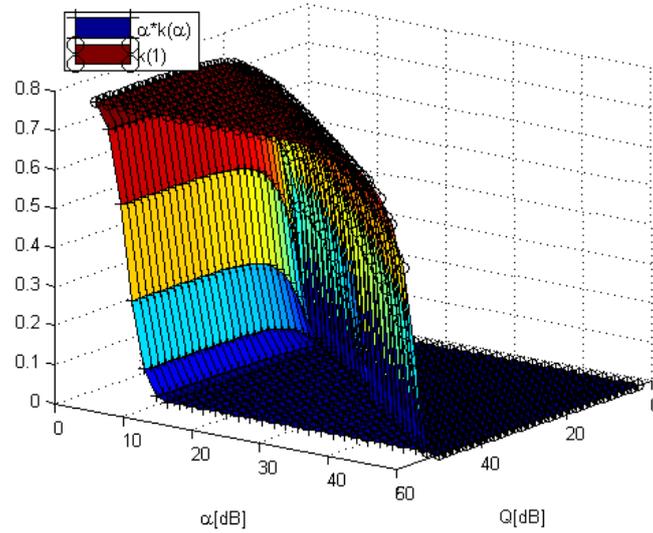

Fig. 8. Graphical verification of the throughput inequality as a function of $\alpha, Q$ (32). The upper curve is $k(1)$, the lower is $\alpha k(\alpha)$ with $k(\alpha)$ defined in (33).